# Ultrahigh Doping of Graphene Using Flame-Deposited MoO$_3$

Sam Vaziri, Victoria Chen, Lili Cai, Yue Jiang, Michelle Chen, Ryan Grady, Xiaolin Zheng, Eric Pop

*Abstract*— The expected high performance of graphene-based electronics is often hindered by lack of adequate doping, which causes low carrier density and large sheet resistance. Many reported graphene doping schemes also suffer from instability or incompatibility with existing semiconductor processing. Here we report ultrahigh and stable *p*-type doping up to ~7×10$^{13}$ cm$^{-2}$ (~2×10$^{21}$ cm$^{-3}$) of monolayer graphene grown by chemical vapor deposition. This is achieved by direct polycrystalline MoO$_3$ growth on graphene using a rapid flame synthesis technique. With this approach, the metal-graphene contact resistance for holes is reduced to ~200 Ω·μm. We also demonstrate that flame-deposited MoO$_3$ provides over 5× higher doping of graphene, as well as superior thermal and long-term stability, compared to electron-beam deposited MoO$_3$.

*Index Terms*— Graphene, Doping, Contact resistance.

## I. INTRODUCTION

Graphene is a two-dimensional (2D) material with unique physics and intrinsic properties [1]. This has resulted in many reports on proof-of-concept electronic and optoelectronic devices with functionalities that cannot be achieved by conventional bulk materials [2-4]. However, many such devices still suffer from issues such as low performance, poor reliability, device-to-device variation, and lack of reproducibility [5-7]. This partly occurs because atomically thin graphene (~0.335 nm monolayer) is very sensitive to perturbations introduced during or after its integration with other materials [8]. This high sensitivity also makes it challenging to achieve effective, reliable, and uniform doping of graphene without sacrificing its electronic quality [9]. As a result, applying conventional semiconductor doping methods (e.g. substituting C atoms with dopants) is very challenging for graphene and other 2D materials.

A more promising route to control the doping of graphene and other 2D materials is to introduce charge carriers from the outside, either with a field-effect (electrostatically or with external charge dipoles) or by external dopants (similar to modulation doping in conventional semiconductors) [10-13]. For example, charge transfer from external metal-oxide layers is one of the most promising mechanisms, being compatible with

This work was supported by Bosch via the Stanford SystemX Alliance. S.V. also acknowledges partial support from a Wallenberg Fellowship. RWG acknowledges support from the NSF GRFP under Grant No. DGE-1656518.
Sam Vaziri, Victoria Chen, Ryan Grady, and Eric Pop are with the Department of Electrical Engineering, Stanford University, Stanford, CA 94305, USA. (e-mail: epop@stanford.edu)
Lili Cai, Yue Jiang, Xiaolin Zheng are with the Dept. of Mechanical Engineering, Stanford University, Stanford, CA 94305, USA.
Michelle Chen and Eric Pop are with the Dept. of Materials Science & Engineering, Stanford University, Stanford, CA 94305, USA.

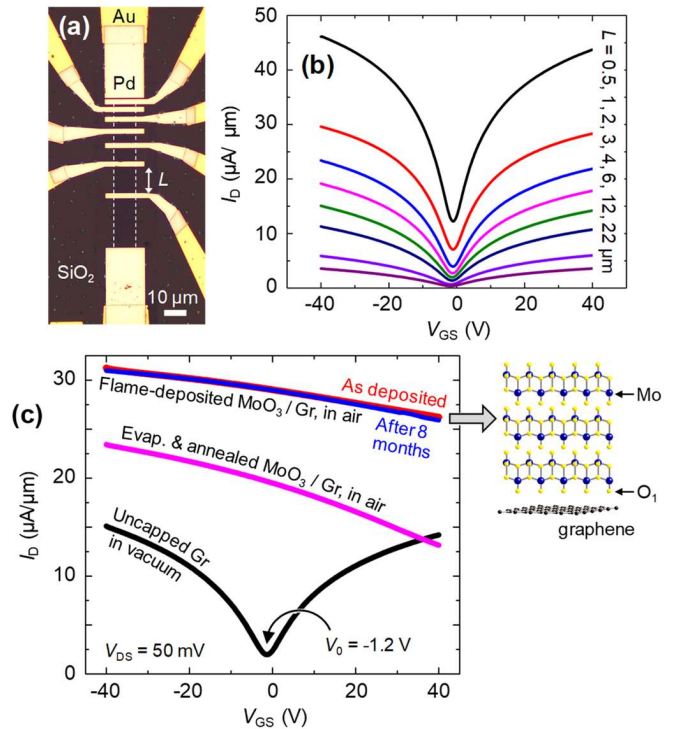

**Fig. 1. Transfer characteristics. (a)** Optical image of a TLM test structure with channel lengths of 0.5 μm, 1 μm, 2 μm, 3 μm, 4 μm, 6 μm, 12 μm and 22 μm. **(b)** Measured current ($I_D$) vs. gate voltage ($V_{GS}$) of uncapped graphene channels in the TLM. **(c)** Measured $I_D$-$V_{GS}$ of a graphene (Gr) channel with $L$ = 4 μm: uncapped (black line), capped with evaporated (magenta line) and flame deposited (red line) MoO$_3$. The blue line shows the same device as in red, after 8 months. The schematic shows the crystalline MoO$_3$ interface with graphene.

semiconductor device processing while minimally degrading the electronic properties of graphene [14, 15]. In this doping scheme, a metal-oxide with a different electron affinity is brought into contact with graphene, transferring charge carriers across the interface and forming interface dipoles. For example, *p*-type doping of graphene can be achieved by capping with molybdenum oxide (MoO$_x$ where $x$ ~ 3) with large electron affinity > 6 eV [10]. MoO$_3$ also has a sufficiently large band gap ($E_G$ ~ 3.7 eV) to be compatible with transparent electronic applications. However, due to varying material quality and measurement conditions, reported results have not been consistent [10, 15-17]. Thus, the effectiveness of graphene doping remains challenging, while little is known about long-term stability [13].

In this letter, we apply MoO$_3$ using a rapid flame vapor deposition technique [18] to effectively induce ultrahigh and stable *p*-type doping of graphene monolayers grown by chemical vapor deposition (CVD). We use electrical measurements, Raman spectroscopy, and modeling to show that a thin layer of ~5 nm flame-deposited MoO$_3$ on monolayer graphene can induce a hole density of ~7×10$^{13}$ cm$^{-2}$ and a metal-graphene contact resistance of ~200 Ω·μm. We compare the effectiveness of this method to the conventional approach of electron-beam (e-beam) evaporated MoO$_3$. The results are averaged over tens of test structures and show long-term stability.

## II. GRAPHENE FIELD-EFFECT TEST STRUCTURES

Monolayer graphene was grown on Cu foil using CVD, and subsequently transferred [19] onto 90 nm SiO$_2$ on $p^+$ Si substrates which also serve as back-gates. After defining graphene channels by photolithography and oxygen plasma, 50 nm-thick Pd contacts were formed using e-beam lithography, e-beam deposition, and lift-off. Similar steps were repeated to form Ti(3 nm)/Au(70 nm) electrical contact pads. Figure 1(a) shows a representative TLM test device with 8 channel lengths, 500 nm ≤ $L$ ≤ 22 μm. Then, the samples were annealed at 250°C in vacuum (5×10$^{-5}$ Torr). Figure 1(b) shows the transfer characteristics of the back-gated graphene field effect transistors (GFETs) in vacuum. These display very symmetric ambipolar behavior with Dirac voltage ($V_0$) near zero and minimal hysteresis, indicating clean, nearly intrinsic graphene channels.

In previous work [18] we showed that atomically thin layers of MoO$_3$ can be grown on layered materials such as WSe$_2$ and graphene, using flame synthesis by van der Waals epitaxy. Unlike e-beam evaporation, the flame-deposited MoO$_3$ on layered materials shows a high degree of crystallinity. Here, we use this method to deposit thin layers (~5 to 10 nm) of MoO$_3$ on the graphene channels to induce a strong $p$-type doping effect.

Figure 1(c) shows measurements of an uncapped GFET in vacuum (black line) which exhibits a maximum hole mobility ~2700 cm$^2$V$^{-1}$s$^{-1}$ and an on/off current ratio of ~7.5 with a hysteresis $|\Delta V_0| \approx 0.5$ V (not shown). After the flame deposition of MoO$_3$, the transfer characteristics of the same device show a significant positive shift of the Dirac voltage (beyond the $V_{GS}$ range we can apply safely without damaging the back-gate oxide), indicative of strong $p$-type doping [Fig. 1(c), red line].

Importantly, re-measuring the same device after 8 months [Fig. 1(c), blue line] shows no significant variation or degradation in the drain current, indicating excellent long-term stability. We attribute this stability to the high crystallinity and stoichiometry of the flame-deposited MoO$_3$ on graphene [18]. For comparison, e-beam evaporated MoO$_3$ was applied on separate samples, which were then annealed in air at 200 °C for 30 min. The effect of annealing is discussed in the next section. For the same channel length of 4 μm, the GFET capped with the evaporated MoO$_3$ [Fig. 1(c), magenta curve] has lower drain current than GFETs capped with the flame-deposited MoO$_3$.

## III. SHEET AND CONTACT RESISTANCE

After process development, we performed six MoO$_3$ flame depositions and two MoO$_3$ evaporation runs. Then we characterized on average twelve TLM structures [20] for each deposition run, with Fig. 2 showing extracted contact resistance $R_C$ and sheet resistance $R_{sh}$ for the uncapped and MoO$_3$-capped samples. The inset shows the TLM plot of resistance vs. length for uncapped devices. In Fig. 2(a), the uncapped devices (blue line) show relatively symmetric Pd-graphene contacts for electrons and holes, suggesting Fermi level pinning near the Dirac point. When approaching the Dirac voltage, $R_C$ increases as the graphene density of states decreases. For holes in doped graphene (red line), we note $R_C \sim 200$ Ω·μm at room temperature, which improves at least by a factor of 2.5 the lowest $R_C$ achieved in the uncapped devices. This contact resistance is comparable to the lowest reported values for Pd with exfoliated monolayer graphene [21].

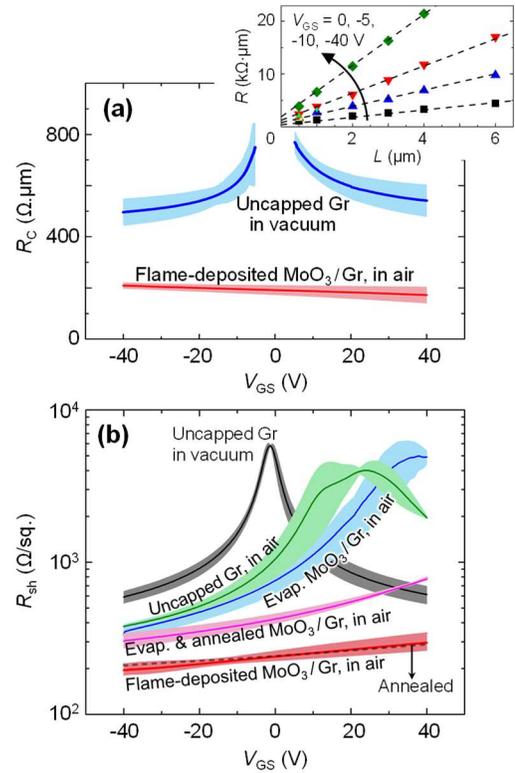

Fig. 2. Contact and sheet resistance from TLM. (a) Pd-graphene (Gr) contact resistance $R_C$ vs. $V_{GS}$ with Gr channel uncapped (blue) and capped with flame-grown MoO$_3$ (red). Inset shows TLM plot and fitting (dashed lines). (b) Extracted graphene sheet resistance for uncapped and capped graphene channels before and after annealing. Solid lines show the average resistance and the shaded regions put bounds on the maximum and minimum values over all the measured TLMs. Black dashes show the average sheet resistance of the flame-deposited MoO$_3$ after annealing.

Figure 2(b) shows the extracted $R_{sh}$ from TLMs for all the uncapped and capped graphene channels before and after annealing. Interestingly, while the annealing changes the doping of graphene capped with evaporated MoO$_x$ [Fig. 2(b), blue and magenta lines], it does not affect the graphene capped with flame-deposited MoO$_3$ [Fig. 2(b), black dashed line]. This shows that the flame-deposited MoO$_3$ is thermally more stable than the evaporated one due to its stoichiometric and crystalline characteristics [18]. Figure 2(b) shows that while electrostatic gating can reduce uncapped graphene sheet resistance by about one order of magnitude, the MoO$_3$ can further and significantly dope graphene, even beyond the electrostatic gating limit. While the large ranges for uncapped samples and those capped with evaporated MoO$_3$ are due to larger hysteresis and device-to-device variation, the annealed graphene devices capped with flame-deposited MoO$_3$ show significantly lower variation.

We attribute the higher doping to two underlying causes. The flame deposition method generates highly crystalline, stoichiometric MoO$_3$, with more terminal O$_1$ sites that create a dipole layer to increase its work function [18, 22]. In addition, flame-deposited MoO$_3$ is horizontally oriented with the (010) planes parallel to the graphene basal plane [Fig. 1(c)]. Because the terminal O$_1$ sites are located on the (010) plane, this orientation creates a stronger dipole layer on the surface, which enables higher charge transfer doping of graphene. In contrast, thermal evaporation generates amorphous, sub-stoichiometric and randomly oriented MoO$_x$ ($x < 3$).

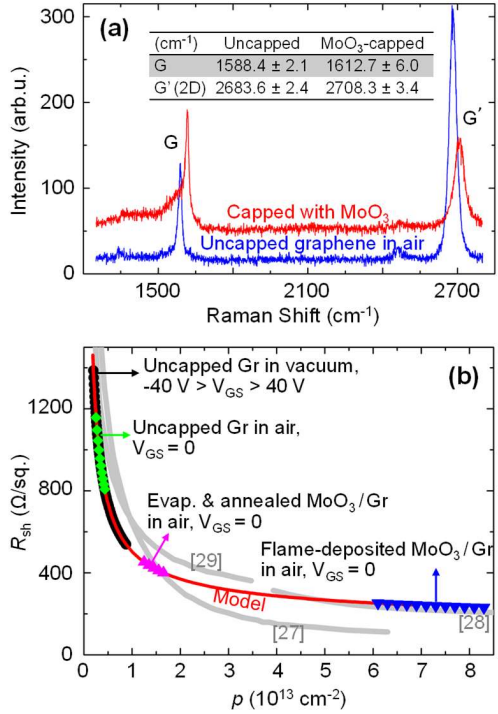

**Fig. 3. Raman analysis and charge carrier extraction.** (a) Raman spectra of the uncapped and capped with flame-deposited MoO$_3$ graphene in air. The inset table shows the average (15 spots each) G and G' peak positions before and after MoO$_3$ deposition. (b) Data fitting to the model and carrier density extraction for capped devices. Gray lines show experimental data from Refs. 27-29, for comparison.

Using Raman spectroscopy, we further monitored the carrier density through the G and G' (also known as 2D) peaks [23]. Figure 3(a) shows the Raman spectra of graphene before (blue line) and after (red line) the flame deposition of MoO$_3$, measured in air. No significant disorder-related D peak (at 1350 cm$^{-1}$) indicates good quality of graphene even after MoO$_3$ deposition. The capped graphene shows a significant blue shift of G and G' (2D) peaks from ~1588.4 cm$^{-1}$ to ~1612.7 cm$^{-1}$, and from ~2683.6 cm$^{-1}$ to ~2708.3 cm$^{-1}$, respectively [inset of Fig. 3(a)]. The G' peak of the capped graphene in Fig. 3(a) exhibits substantial intensity drop and line broadening, which are mainly ascribed to the large hole doping [23]. Comparing these peak shifts and the intensity ratio of G and G' peaks for the capped graphene [I(G')/I(G) < 0.77] to the electrochemical-gate calibrated graphene in Ref. 23 suggests $p$-doping >3×10$^{13}$ cm$^{-2}$ in the graphene capped with flame-deposited MoO$_3$. However, we note that the possible strain effect of the capping layer can contribute to the peak shifts and variations affecting the accuracy of the carrier density extraction from the Raman spectra.

Because the Dirac voltage is outside the measurable $V_{GS}$ range for the capped devices, it is not possible to directly extract the carrier density from the measured $I_D$ vs. $V_{GS}$. Hence, we applied a well-tested model [24, 25] to estimate the carrier density in the graphene channels from our electrical results. This model is based on the power law behavior of mobility vs. carrier concentration $\mu(p) \propto p^{-\alpha}$ [26]. We fit our uncapped graphene data [Fig. 3(b), thick black line] as $R_{sh} = [1 + (p/p_{ref})^{\alpha}]/(qp\mu_0)$, shown with the red line, where $\mu_0 = 3350$ cm$^2$V$^{-1}$s$^{-1}$, $p_{ref} = 5.14 \times 10^{12}$ cm$^{-2}$, and $\alpha = 0.8$. $R_{sh}$, $p$ and $q$ are sheet resistance, hole density, and elementary charge, respectively. Then the carrier densities in the capped devices with evaporated [magenta symbols in Fig. 3(b)] and flame-deposited MoO$_3$ [blue symbols in Fig. 3(b)] are extracted. For comparison, the gray lines in Fig. 3(b) show experimental data from previous Hall measurements at high carrier densities achieved by electrolyte gating [27, 28] and nitric acid doping [29], demonstrating similar behavior. Table 1 summarizes our average sheet resistance, estimated carrier densities, and the resistance ratio ($R/R_0$) at zero gate voltage, where $R_0$ and $R$ are the resistances of uncapped and capped-with-MoO$_3$ graphene channels, respectively.

**Table 1.** Sheet resistance, charge carrier density, and resistance ratio averaged over all the measured graphene devices.

|  | $R_{sh}$ (Ω/sq.) | $p$ (cm$^{-2}$) | $R/R_0$ (vac.) | $R/R_0$ (air) |
|---|---|---|---|---|
| Gr in vac. | 5700 | - | - | - |
| Gr in air | 980 | 3.4×10$^{12}$ | 17% | - |
| Gr / evap. MoO$_3$ | 423 | 1.4×10$^{13}$ | 7% | 43% |
| Gr / flame MoO$_3$ | 230 | 7.2×10$^{13}$ | 4% | 23% |

## IV. CONCLUSION

In conclusion, we applied our flame vapor deposition technique to grow thin films of stoichiometric crystalline MoO$_3$ on monolayer CVD graphene. The resulting structures exhibit an ultrahigh and stable $p$-type charge transfer doping up to ~7×10$^{13}$ cm$^{-2}$, five times higher than control samples with e-beam evaporated MoO$_3$. Using this approach, the metal (Pd)-graphene contact resistance was also reduced by a factor of 2.5 to ~200 Ω·µm. Our fabrication process is CMOS-compatible and can be used to realize competitive performance in graphene electron devices, as well as low-sheet-resistance transparent conductors.